\documentclass[floatfix,twocolumn,showpacs,preprintnumbers,amsmath,amssymb,pra,superscriptaddress,longbibliography]{revtex4-1}

\usepackage{graphicx}
\usepackage{dcolumn}
\usepackage{bm}
\usepackage{float} 
\usepackage{lipsum}
\usepackage{enumitem,amssymb} 
\usepackage{color}
\usepackage[usenames,dvipsnames,svgnames,table]{xcolor}
\usepackage{hyperref}
\hypersetup{
    colorlinks = true,
    citecolor = blue
}
\newlist{todolist}{itemize}{2}
\setlist[todolist]{label=$\square$}
\usepackage{amsmath} 

\usepackage{color}

\usepackage{siunitx}
\usepackage{pifont}

\usepackage{leftidx}
\usepackage{tabulary}
\usepackage{tabularx,booktabs}
\usepackage[capitalise]{cleveref}
\usepackage[utf8]{inputenc}
\usepackage[T1]{fontenc}
\usepackage{csquotes} 

\graphicspath{{Figures/}}

\newcommand{\casus}{%
    Center for Advanced Systems Understanding (CASUS),  
    D-02826 G\"orlitz, 
    Germany}

\newcommand{\hzdr}{%
    Helmholtz-Zentrum Dresden-Rossendorf (HZDR), 
    01328 Dresden, 
    Germany}

\newcommand{\snl}{%
    Computational Multiscale Department, 
    Sandia National Laboratories, 
    87185 Albuquerque, NM, 
    United States}
    
\newcommand{\cea}{%
    CEA, DES, IRESNE, DEC, SESC, LM2C, 
    F-13108 Saint-Paul-Lez-Durance, France}
    
\begin{document}


\title{Transferable Interatomic Potentials for Aluminum from Ambient Conditions to Warm Dense Matter} 

\author{Sandeep Kumar} 
\email{s.kumar@hzdr.de} 
\affiliation{\casus}
\affiliation{\hzdr}

\author{Hossein Tahmasbi} 
\affiliation{\casus}
\affiliation{\hzdr}

\author{Kushal Ramakrishna} 
\affiliation{\casus}
\affiliation{\hzdr}

\author{Mani Lokamani} 
\affiliation{\hzdr}

\author{Svetoslav Nikolov}
\affiliation{\snl}

\author{Julien Tranchida}
\affiliation{\cea}

\author{Mitchell A. Wood}
\affiliation{\snl}

\author{Attila Cangi} 
\email{a.cangi@hzdr.de}
\affiliation{\casus}
\affiliation{\hzdr}

\date{\today}
             
\begin{abstract} 
We present a study on the transport and materials properties of aluminum spanning from ambient to warm dense matter conditions using a machine-learned interatomic potential (ML-IAP). Prior research has utilized ML-IAPs to simulate phenomena in warm dense matter, but these potentials have often been calibrated for a narrow range of temperature and pressures.
In contrast, we train a single ML-IAP over a wide range of temperatures, using density functional theory molecular dynamics (DFT-MD) data.
Our approach overcomes computational limitations of DFT-MD simulations, enabling us to study transport and materials properties of matter at higher temperatures and longer time scales. We demonstrate the ML-IAP transferability across a wide range of temperatures using molecular-dynamics (MD) by examining the thermal conductivity, diffusion coefficient, viscosity, sound velocity, and ion-ion structure factor of aluminum up to about 60,000K, where we find good agreement with previous theoretical data.
\end{abstract}    
 
\maketitle


\section{Introduction}  
Warm dense matter (WDM) is a state of matter characterized by densities ranging from solid density to a few orders higher than solid density and temperatures ranging from a few eV to a few keV~\cite{Graziani_14_book}. It is defined by two parameters: the ionic coupling parameter $\Gamma$, which is the ratio of the average kinetic energy to the average potential energy, and the electron degeneracy parameter $\Theta$, which is the ratio of the average kinetic energy to the average Fermi energy. WDM is defined as the region of parameter space where $\Gamma \approx 1$ and $\Theta \approx 1$~\cite{Riley_2000, Ma_13, Bonitz_20,dornheim2023electronic}.

The study of WDM is important for advancing our understanding of phenomena that both occur in nature and are generated in the laboratory. In nature, such conditions can be found in the cores of giant planets and exoplanets~\cite{Guillot_99}. In the laboratory, WDM is generated at various facilities around the world, including pulsed power facilities such as the Z-Machine~\cite{Sinars_20} and bright photon sources such as the European X-Ray Free-Electron Laser Facility (European XFEL)~\cite{Voigt_21}, the Linac Coherent Light Source (LCLS)~\cite{Fletcher_15, McBride_18, frydrych2020demonstration}, and the National Ignition Facility (NIF)~\cite{Haynam_07}. Inertial confinement fusion experiments also involve the creation of WDM when fuel capsules are heated towards ignition~\cite{Betti_16,Graziani_14}.

Modeling and simulating WDM is challenging due to its unique properties, which fall in a regime that is too dense for standard plasma theories and too hot for condensed matter theories to be applicable. To study the dynamical and thermodynamical properties of WDM, density functional theory molecular dynamics (DFT-MD) simulations have been widely used~\cite{Desjarlais_02, Desjarlais_03, Mazevet_05, Nettelmann_08, PhysRevB.103.125118, PhysRevLett.125.235001, French_09}. However, there are several limitations to the use of DFT-MD for this purpose. First, DFT-MD becomes computationally infeasible at higher temperatures, making it difficult to study the properties of WDM under these conditions. Second, finite-size effects, which are caused by the limited number of atoms that can be simulated on current high-performance computing (HPC) platforms (typically a few hundred atoms), can lead to inaccurate results for many observables. Finally, DFT-MD simulations are typically limited to time scales of around 100 ps, making it difficult to accurately study long-timescale phenomena such as ionic transport.

In the past, embedded atom models (EAMs)~\cite{Daw_83,Daw_84, Foiles_86} have been proposed as potentials for large-scale MD simulations of WDM~\cite{Norman_13,Chen_18}, but they have been found to be inconsistent with DFT-MD simulations over a wide range of temperatures and pressures. On the other hand, the use of Yukawa pair potentials (shielded Coulomb potentials) has been reported in the literature, but these approaches are not sufficiently accurate in the strong coupling regime due to the absence of mean field effects~\cite{Vorberger_12, White_13, Dharma_22}.

Recently, MD simulations using machine learning-based interatomic potentials (ML-IAPs) have been shown to overcome these computational limitations while maintaining the accuracy of DFT-MD simulations~\cite{Behler_07,Bartok_10,Thompson_15, Zhang_18,Behler_16, Tisi_21, nikolov2021data}. These ML-IAPs have the potential to enable the study of thermodynamic properties in WDM at higher temperatures and longer time scales~\cite{Cheng_21, Liu_21, Stanek_21, Schoner_22, Schorner_22_Cu, Willman_22}, enabling a deeper understanding of this important state of matter. The materials properties the ML-IAP enables to simulate via large-scale MD simulations are important for understanding the dynamics of planetary interiors and inertial confinement fusion plasmas. For example, the diffusion of particles plays a significant role in fuel degradation during inertial confinement fusion~\cite{Murphy_16}, while the thermal conductivity of materials is an important quantity for understanding the cooling process of a planet's core~\cite{Buffett_02,Sandeep_21,Wieland_22}.

In this study, we use ML-IAPs to simulate WDM over a wide range of temperatures. Previous research has used ML-IAPs for WDM simulations, but these potentials have often been calibrated for specific temperature and pressure ranges~\cite{Liu_21,Schoner_22,Schorner_22_Cu,nikolov2022dissociating}. In contrast, we train a single ML-IAP that can be used across a range of temperatures using DFT-MD data. To generate the ML-IAP, we use the Spectral Neighbor Analysis Potential (SNAP) method~\cite{Thompson_15}, which represents the local environment of each atom through bispectrum components of the local atomic density projected onto a basis of hyperspherical harmonics in four dimensions. These components serve as descriptors in the SNAP method. We use the DAKOTA optimization software~\cite{adams_20} to tune the ML-IAP's hyperparameters, such as the cutoff distance for the potential and the weight of specific training data sets. By using this approach, we are able to accurately simulate WDM over a wide range of temperatures. We focus our study on aluminum which has been subject to various experimental measurements~\cite{Mckelvey_17, Ma_13, Nagler_09, Leguay_13, Mo_17, Sperling_15, Witte_17}.

We center our investigation on assessing the ML-IAP's ability to extrapolate to higher temperatures beyond the range of the training data. In order to evaluate the accuracy of our ML-IAP, we conduct large-scale MD simulations to calculate various properties over a range of temperatures from 300~K (0.0259~eV) to 58022~K (5~eV), including the thermal conductivity, viscosity, diffusion coefficient, sound velocity, and ion-ion structure factor. In the diffusion coefficient calculation we extend the mentioned temperature range further up to 116040~K (10~eV). 

Our paper is organized as follows. In~\cref{sec:mthd}, we describe the methods used to train the ML-IAPs and to calculate the quantities using MD simulations. In~\cref{sec:rlt}, we present the results of our investigation on the temperature transferability of the ML-IAP by calculating various transport and material properties. Finally, in~\cref{sec:concl}, we provide our conclusion.

\section{Methods}
\label{sec:mthd}

\subsection{Generation of the ML-IAP}
We generate an ML-IAP based on the SNAP methodology~\cite{Thompson_15,Wood2018} for aluminum in a large temperature range. Training data is generated using DFT-MD calculations at ambient mass density (2.7 $\text{g/cm}^3$) using simulation cells containing 108 and 256 atoms over a temperature range of 300 to 10000 K. These simulations are performed using the VASP software package~\cite{Kresse_96,Kresse_96_CMS,Kresse_99} with Kohn-Sham orbitals expanded in plane waves and PAW pseudopotentials used for the electron-ion interaction~\cite{Hohenberg_64,Kohn_65,Mermin_65} with a core radius of $r_{c}$ = 1.9~$a_{B}$ where $a_{B}$ is the Bohr radius. The plane wave cutoff is set to 450~eV and the convergence in the total energy in each self-consistency cycle is set to $10^{-5}$. The PBE exchange-correlation functional~\cite{Perdew_96} is applied throughout. The DFT-MD simulations are run for 10000 steps with a time step of 0.2 fs and are thermostated using the Nose-Hoover method~\cite{Nose_84,Hoover_85}. We investigated the time evolution of the energy and characteristics of the radial distribution function to test the equilibration of the system and found that the system gets equilibrated around 10000 time steps. We sample the Brillouin zone on a 2$\times$2$\times$2 grid of $\emph{k}$-points and gamma-point for the solid and liquid configurations, respectively, and increase the number of bands in the simulations as the temperature increases. The number of bands is varied from 324 at the minimum temperature (300~K) for a supercell containing 108 atoms to 896 at the maximum temperature (10000~K) considered in the data set for a supercell containing 256 atoms. By doing so, we achieve convergence in the total energy in each self-consistency cycle up to $10^{-5}$.

In the next step, we create an ML-IAP based on this training data. Following the SNAP methodology, the total energy of the system of N atoms with positions $\mathbf{r}^N$ is decomposed into 
\begin{equation}
E (\mathbf{r}^{N}) = E_\mathrm{ref} (\mathbf{r}^{N}) + \sum_{i=1}^N E_\mathrm{SNAP}^i\,,
\end{equation}
where $E_\mathrm{ref}$ denotes a reference potential energy and $E_\mathrm{SNAP}^i$ the total energy of atom $i$ relative to the atoms in its neighborhood. 
Here, the presence of the reference potential can improve the accuracy of the potential energy which need not be parametrized by the training data but can be included from known limiting cases. In this regard, we use the Ziegler-Biersack-Littmark (ZBL) potential~\cite{Ziegler_85} as a reference potential. The ZBL model for aluminum is derived from ion stopping power data, thus providing an accurate description of short-range (higher-energy) atomic interactions. 
The SNAP ML-IAP energy of each atom is expressed as
\begin{equation}
E_\mathrm{SNAP}^i =  \boldsymbol{\beta} \cdot  \mathbf{B}^i
\end{equation}
in terms of a linear combination of bispectrum components $\mathbf{B}^i$ and linear coefficients $\beta_k$.
Consequently, the forces on an atom $j$ are calculated from the derivatives of the bispectrum components of atom $i$ with respect to changes in the positions $\mathbf{r}^j$ which reads
\begin{equation}
F_\mathrm{SNAP}^j  = -\nabla_j  \sum_{i=1}^N E_\mathrm{SNAP}^i = -\boldsymbol{\beta} \cdot \sum_{i=1} ^{N} \frac{\partial \mathbf{B}^i}{\partial \mathbf{r}^j} \;. 
\end{equation}
The bispectrum components $\mathbf{B}^i$ represent the atomic density of those atoms in the neighborhood of atom $i$ with a cutoff distance $R_\mathrm{cut}$. They are related to 4D hyperspherical harmonics and do preserve invariance under symmetry operations~\cite{Wood2018}.
The trainable parameters, $\boldsymbol{\beta}$, of the SNAP ML-IAP are determined by weighted linear regression in an iterative process to reproduce the total energy and atomic forces in each atomic configuration of the DFT-MD training data set. The linear regression is carried out using the FitSNAP software package~\cite{fitsnap,rohskopf2023fitsnap}.
To this end, we pre-process the DFT-MD data by extracting the total energies and the forces on all the atoms. The training datasets and parameters are listed in Tab.~S1~\cite{sup_mat}. The DFT-MD data is grouped into intervals of 100 K in the temperature range from 300 to 1000 K, and at intervals of 1000 K from 1000 to 10000 K. For each temperature, we randomly select approximately 100 snapshots of equilibrated atomic configurations. We include low-temperature data from 300-900 K in the training because a diverse dataset improves the quality of the trained ML-IAP. Based on this data, we generate a SNAP ML-IAP over the desired temperature range.

Finally, we need to determine the hyperparameters of the ML-IAP. These include the cutoff radius $R_\mathrm{cut}$, which defines the length scale of the atomic environment considered, $J_\mathrm{max}$, which determines the number of terms in the expansion of the atomic density in terms of bispectrum components, and the weights of the energy and force training sets. To fix the number of terms in the expansion, we set $2 J_\mathrm{max}=6$, and used the DAKOTA software package~\cite{adams_20} to optimize the remaining parameters. The cutoff radius was allowed to vary from 4.8 to 5.8 \AA, whereas both the energy weights (EW) and force weights (FW) were optimized within the range from 0.5/(\textit{number of snapshots}) to 1.5/(\textit{number of snapshots}). The DAKOTA iterations were run approximately 1300 times to find the optimal parameters, resulting in a cutoff radius of 5.323 \AA{} and the energy and force weights listed in Tab.~S1~\cite{sup_mat}. 

To assess the accuracy of the SNAP ML-IAP, we compare its predictions of energy and force with the reference DFT-MD values. This comparison is displayed in Fig.~S2~\cite{sup_mat} as a correlation plot, where the ideal result is a straight line (shown in red). The SNAP predictions (represented by blue points) have a certain spread and tilt around this line, which gives a qualitative measure of the error. The mean absolute error (MAE), root mean square error (RMSE), and standard deviation in energy values for the trained ML-IAP were 6.27 meV/atom, 8.7 meV/atom, and 72.56 meV/atom, respectively. For the forces, these errors per atom were 0.22 eV/\AA, 0.32 eV/\AA, and 0.73 eV/\AA, respectively. Histograms displaying the difference between the SNAP prediction and DFT-MD values in the energy and force given in Fig.~S1~\cite{sup_mat}. These errors indicate that the trained ML-IAP is accurate and suitable for use in large-scale MD simulations.

\subsection{Time-averaged materials properties from ML-IAP-driven molecular dynamics simulations}
Large-scale MD simulations are carried out using the open-source LAMMPS~\cite{Plimpton_95} code. A three-dimensional cubic box with aluminum atoms in face centered cubic (fcc) configuration is created. The atoms interact with each other via the SNAP ML-IAP. Periodic boundary conditions are chosen for the simulations. We consider the aluminum atom's mass $m = 26.981539$ u (u is an atomic mass unit), atomic number $Z = 13$, and lattice constant $a = 4.048$ \AA. The lattice constant sets the mass density of the aluminum atoms to its ambient value $\rho = 2.7$ $\text{g/cm}^3$. We chose the simulation time step as 1 fs which ensures a fine discretization along the temporal domain and good resolution of the underlying kinetics.

The thermalization of the system is achieved by evolving positions and velocities from a canonical ensemble (NVT) using a Nose-Hoover~\cite{Nose_84, Hoover_85} thermostat. For each temperature the system is initially evolved in the NVT ensemble for 200 ps to ensure an equilibrium state. After this initial $200$ ps NVT run, the atomic configuration achieves thermodynamic equilibrium with the desired temperature and is ready for the transport calculation run. Transport properties are calculated in the microcanonical ensemble (NVE) on a total time period of 10 ns.

\section{Results}
\label{sec:rlt}
The purpose of our study is to evaluate the applicability of the trained SNAP ML-IAP in a wide temperature range, including the challenging conditions of WDM. To achieve this, we analyze various ionic transport and material properties, such as thermal conductivity, viscosity, diffusion coefficient, sound velocity, and ion-ion structure factor. Our study focuses on aluminum with a fixed mass density of 2.7 $\text{g/cm}^3$ and covers a broad temperature range from 300~K to 58022~K (along an isochore). To assess the accuracy of the SNAP ML-IAP, we test its performance both within and outside the range of the provided training data.

\subsection{Thermal Conductivity}
We compute the lattice contribution to the thermal conductivity $\kappa$ from the well-known Green-Kubo formula~\cite{Green_54, Kubo_57}
\begin{equation}
\kappa = \frac{V}{3 k_B T^2} \int^\infty_0  \langle \mathbf{J}(t) \mathbf{J}(0) \rangle dt \;.
\label{eq:int_j}
\end{equation}
It yields the thermal conductivity from the ensemble average of the auto-correlation $\langle \mathbf{J}(t) \mathbf{J}(0)\rangle$ of the heat flux $J$. Here, $V$ denotes the volume, $k_B$ the Boltzmann constant, and $T$ the temperature. The heat flux is defined as
\begin{equation}
\mathbf{J} = \frac{1}{V}\bigg[ \sum_{i=1} ^{N} e_i \mathbf{v}_i + \frac{1}{2}\sum_{i<j} ^{N} [\mathbf{F}_{ij} (\mathbf{v}_i + \mathbf{v}_j)] \mathbf{r}_{ij} \bigg]\,,
\end{equation}
where $e_i$, $\mathbf{r}_{ij}$, $\textbf{v}_i$, and $\textbf{F}_{ij}$ are the per-atom energy (potential and kinetic), interatomic distance between atoms, velocity, and forces between atoms, respectively. 
The integration in \cref{eq:int_j} is truncated after a sufficiently large time period which is determined by the decay of the auto-correlation function. We truncate it at 6 ps for low temperatures (300~K to 950~K), at 1.5 ps at 1000K, and at 1 ps for higher temperatures (1500~K to 58022~K). Taking a large time interval for integration can introduce additional noise into the thermal conductivity \cite{Humbert_19}. 

\begin{figure}[!hbt]   
\centering       
\includegraphics[width=1.0\columnwidth]{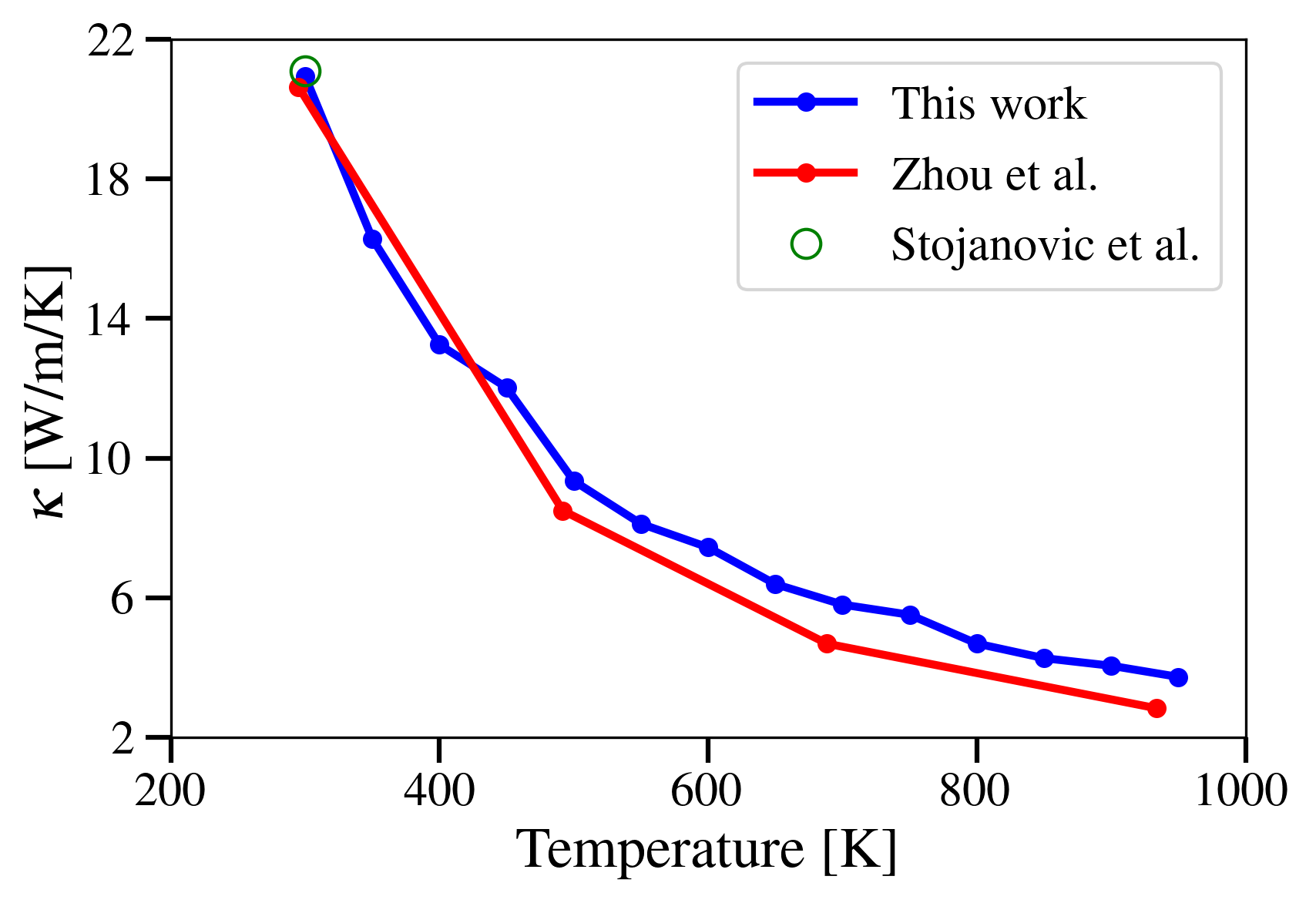}
\caption{Lattice thermal conductivity of aluminum at 2.7 $\text{g/cm}^3$ as a function of temperature from 300~K (0.0259~eV to 950~K (0.0819~eV). We compare our results with the non-equilibrium MD simulation results of~\citet{Zhou_07} and Boltzmann transport equation-based result of~\citet{Stojanovic_10}.}  
\label{fig:thclowT} 
\end{figure}

\begin{figure}[!hbt]   
\centering       
\includegraphics[width=1.0\columnwidth]{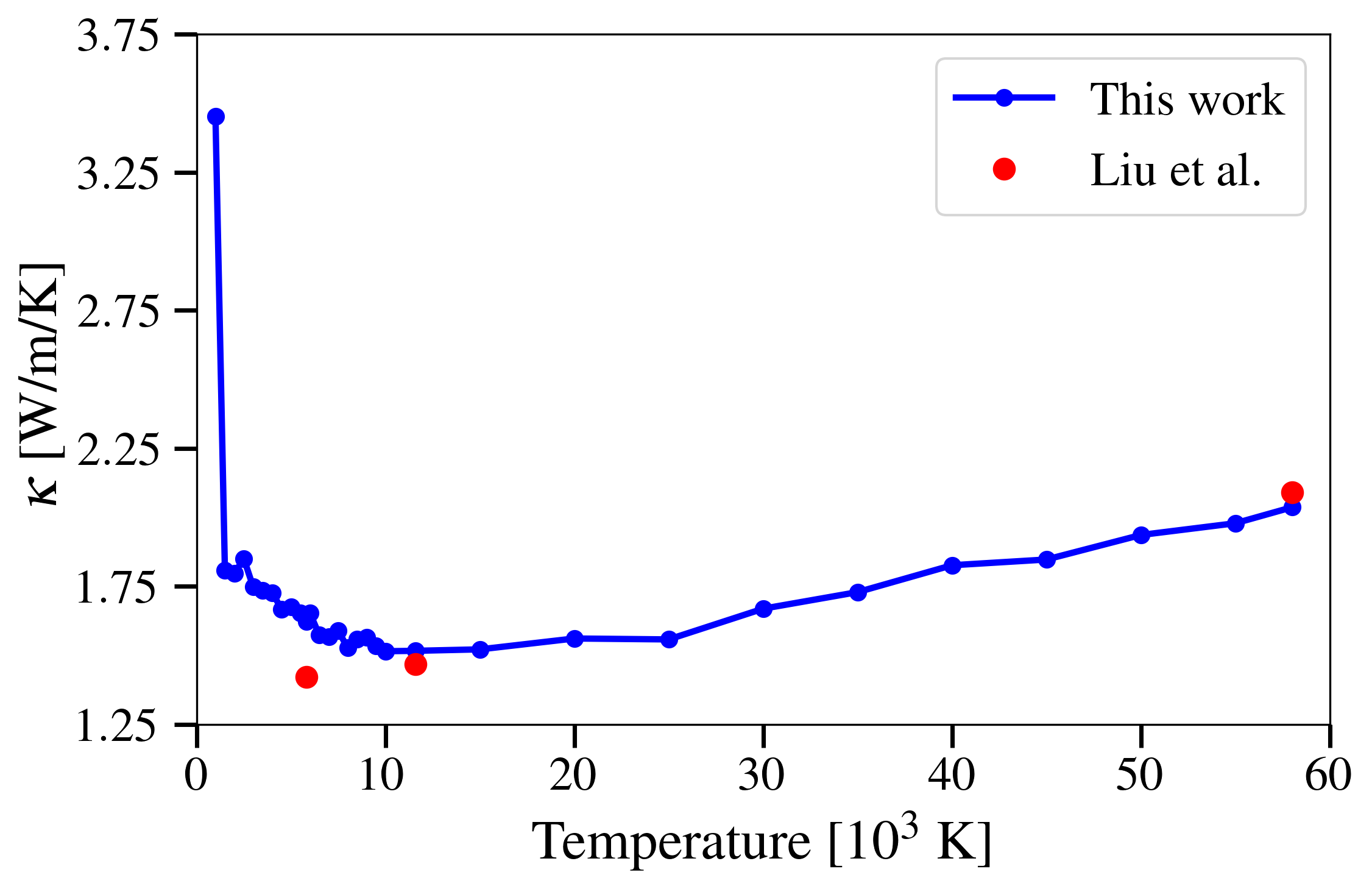}
\caption{Lattice thermal conductivity of aluminum at 2.7 $\text{g/cm}^3$ as a function of temperature from 1000~K (0.0862~eV) to 58022~K (5~eV). The SNAP ML-IAP is trained from 300~K (0.0259~eV) to 10000~K (0.861~eV). We compare our results with the results of~\citet{Liu_21}, which is calculated by the deep potential method trained on each temperature.}  
\label{fig:thchighT} 
\end{figure}

\cref{fig:thclowT} and~\cref{fig:thchighT} display our findings on the temperature-dependent lattice contribution to thermal conductivity. In pure metals, such as aluminum, the thermal conductivity is primarily determined by free electrons, which serve as heat carriers. 

The comparison of our low-temperature predictions with the non-equilibrium MD calculation of~\citet{Zhou_07} and the Boltzmann transport equation calculation of~\citet{Stojanovic_10} are shown in \cref{fig:thclowT}. The excellent agreement illustrates the accuracy of our ML-IAP. The agreement with other predictions at elevated temperatures shows that the ML-IAP accurately reproduces the phonon heat conduction across a large range of solid-phase temperatures.

In the range from 300 K to 10000 K, thermal conductivity decreases with an increase in temperature, but as temperature rises ($>$ 10000~K), the thermal conductivity starts increasing (see~\cref{fig:thclowT} and~\cref{fig:thchighT}). In~\cref{eq:int_j}, there is a kinetic and a potential contribution to the thermal conductivity. The kinetic contribution increases with an increase in temperature and dominates in heat transport at higher temperatures ($>$ 10000~K). The vibrational modes scatter in collision with other modes resulting in a reduced heat transport. The number of vibrational modes decreases with a decrease in the temperature of the crystal, but due to less scattering, it allows the remaining phonons to travel further, which leads to higher thermal conductivity at a lower temperature.

Next, we compare our predictions with those of~\citet{Liu_21} in~\cref{fig:thchighT}. Our analysis indicates that our trained SNAP ML-IAP performs well within the range of the provided training data (300 to 10000 K) and also delivers remarkable accuracy outside this range. However, we observe a significant deviation between our predictions and those of Liu et al. at approximately 5802~K (0.5~eV). This difference may be due to noise in the calculation of~\citet{Liu_21} or the use of orbital-free density functional theory (OFDFT) data in their training. OFDFT is appropriate for high temperatures but may become inaccurate at lower temperatures due to the approximations inherent in the kinetic energy functional that dominates the total energy. Additionally, it is worth noting that~\citet{Liu_21} trained and assessed their ML-IAP at each temperature displayed in their study, whereas we assess the quality of our ML-IAP outside the range of training data. Despite the sensitivity of thermal conductivity to the interatomic potential, our ML-IAP produces reliable outcomes over a wide temperature range.

Note that a long simulation time is required to calculate a noiseless thermal conductivity~\cite{Scheiner_19}. Therefore, we ran these calculations for 10 ns before collecting data for processing. A shorter simulation time is possible, but it will require smoothing of data along the temperature axis to suppress noise. 

We also investigated the effect of finite simulation cell size on the thermal conductivity and found that it is negligible for systems with more than 5000 atoms (see Fig. S3~\cite{sup_mat}).

\subsection{Viscosity}
We also use the Green-Kubo formalism to calculate the viscosity 
\begin{equation}
\eta = \frac{V}{3 k_B T}\int_0^\infty \langle P_{\alpha \beta}(t) P_{\alpha \beta}(0) \rangle dt\ \;.
\end{equation}
Similarly, it is given by an autocorrelation function of the off-diagonal elements of the stress tensor
\begin{equation}
P_{\alpha \beta} = \frac{1}{V}\left[ \sum_{i=1}^{N} \frac{p_i^\alpha p_i^\beta}{m_i} - \sum_{i}^{N} \sum_{i>j}^{N} (\alpha_i - \alpha_j) F_{ij} \right] \,,
\label{eq:vis_eq}
\end{equation}
where $N$ is the total number of atoms, $p_i^\alpha$ and $p_i^\beta$ are the $\alpha$ and $\beta$ components of the momentum of the $i$th atom, $m_i$ is the mass of the $i$th atom, $\alpha_i$ are the Cartesian coordinates of the atoms, $F_{ij}$ are the forces between atoms $i$ and $j$, and $V$ is the volume of the simulation box.

It is important to note that the viscosity calculated using the Green-Kubo formula is dependent on the length of the simulation and that longer simulations are required for accurate results. 

\begin{figure}[!hbt]   
\centering       
\includegraphics[width=1.0\columnwidth]{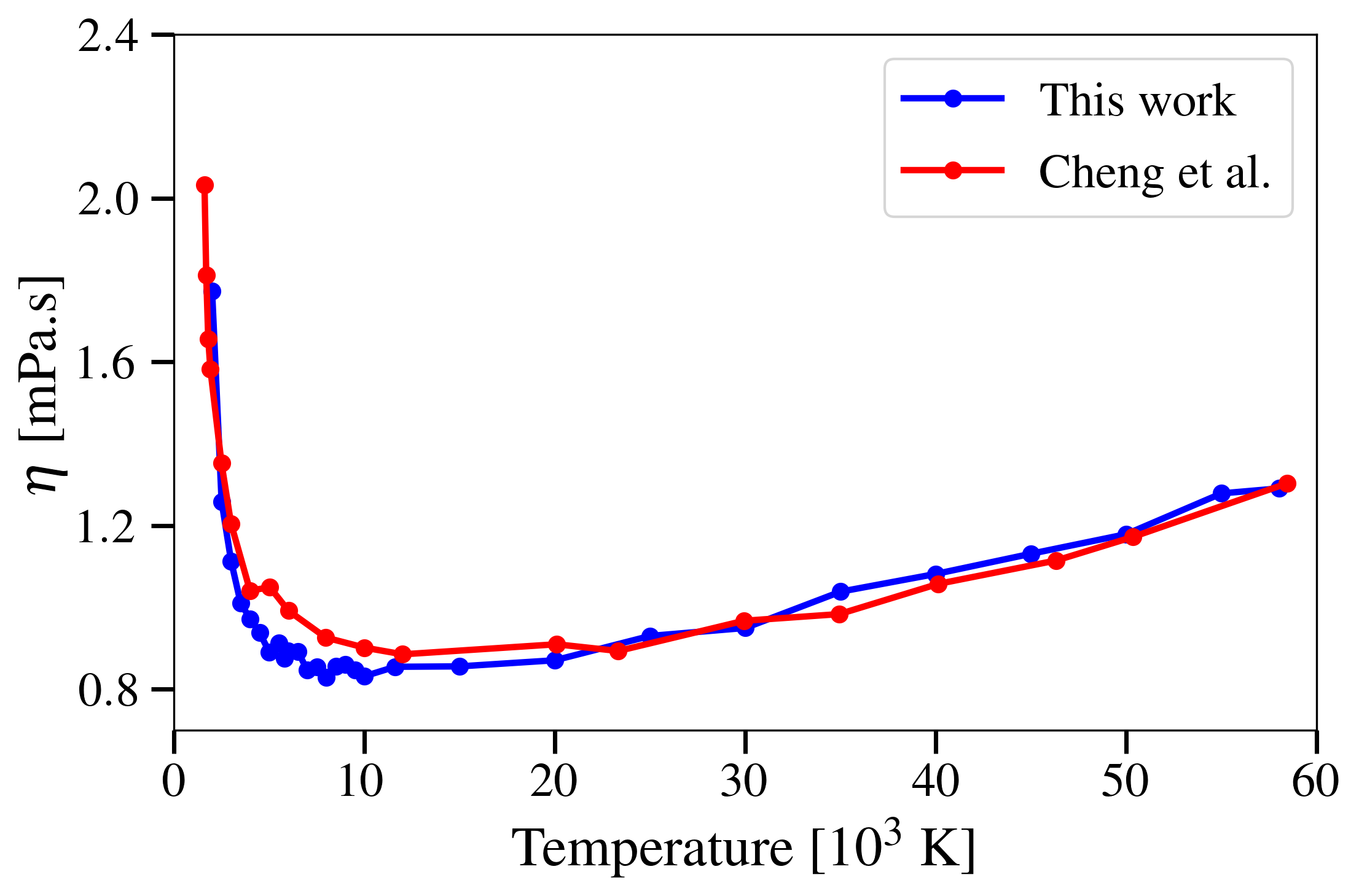}
\caption{Viscosity of aluminum at 2.7 $\text{g/cm}^3$ as a function of temperature from 2000~K (0.1723~eV) to 58022~K (5~eV). Our values are compared with the viscosity values calculated by~\citet{Cheng_21}.}  
\label{fig:vis} 
\end{figure}

\cref{fig:vis} displays the temperature-dependent viscosity of our system. We observe a decreasing trend in viscosity with increasing temperature from 2000 K to 10000 K, followed by a reversal of this trend at temperatures above 10000 K. We compared our results with those obtained with an ML-IAP by~\citet{Cheng_21} and found that our trained ML-IAP accurately predicts the viscosity at the training points, as well as in the intermediate and extrapolation regions.

As with thermal conductivity, the viscosity of our system has both kinetic and potential contributions, as described by~\cref{eq:vis_eq}. At lower temperatures, the momentum transport is dominated by the potential contribution, while at higher temperatures, the kinetic contribution becomes more significant. In the intermediate temperature range, both contributions play a role, resulting in longer relaxation times and higher noise levels in the transport properties, even with simulations as long as 10 ns. Nonetheless, our results provide valuable insights into the temperature-dependent behavior of the viscosity of our system.

\subsection{Diffusion Coefficient}
We also investigate the diffusion coefficient, $D$, which is calculated using the mean squared displacement of atoms:
\begin{equation}
D = \frac{1}{6t} \langle\sum_{i=1} ^{N} [\mathbf{r}_i(t) - \mathbf{r}_i(0)]^2 \rangle \,,
\end{equation}
where $\mathbf{r}$ is the position of the $i$th atom at a given time $t$. The slope of the mean squared displacement versus time plot gives the values of the diffusion coefficient.

\begin{figure}[!hbt]   
\centering       
\includegraphics[width=1.0\columnwidth]{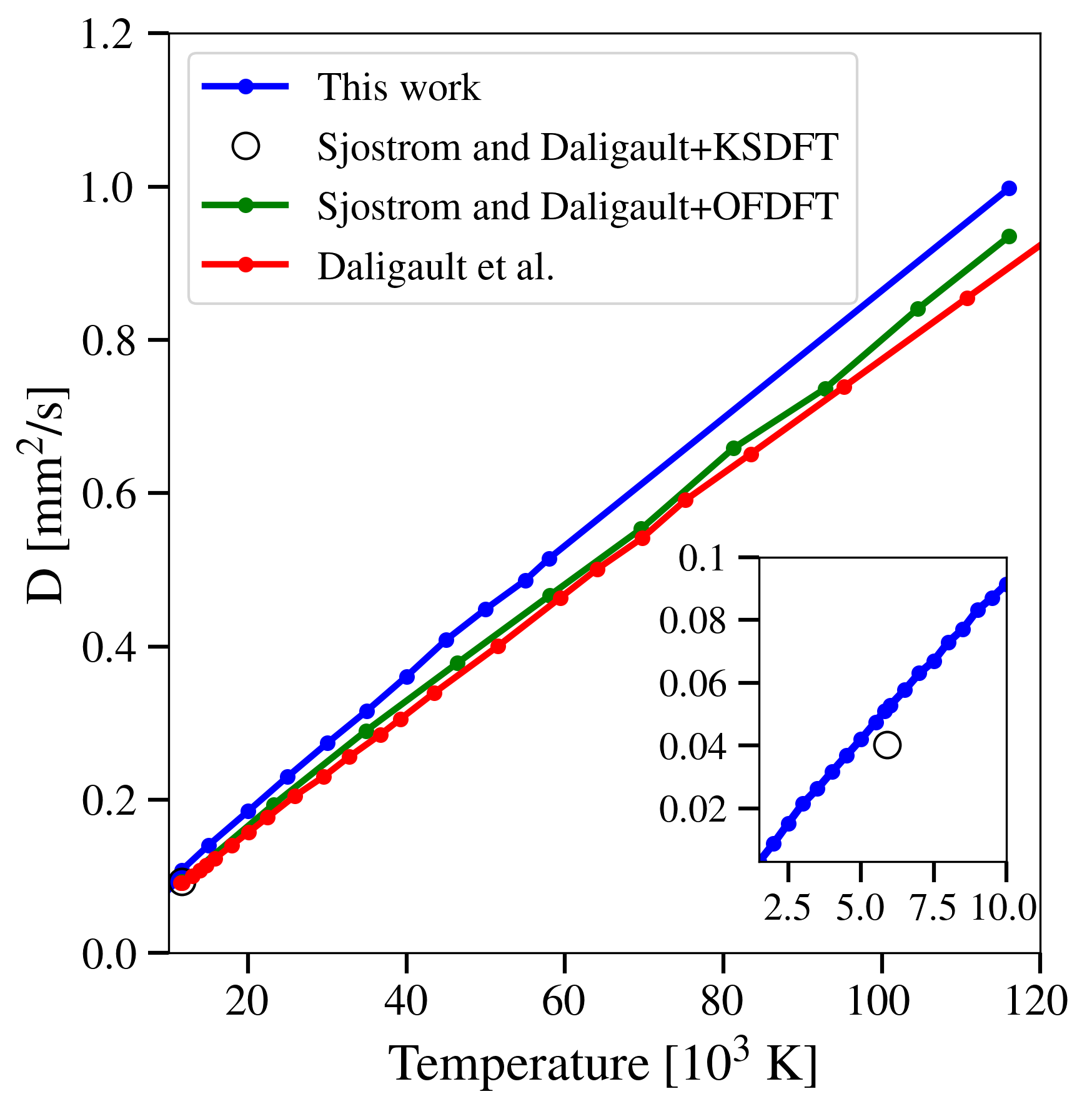}
\caption{Diffusion coefficient of aluminum at 2.7 $\text{g/cm}^3$ as a function of temperature from 1500~K (0.129~eV) to 116040~K (10~eV). We compare our results with the KSDFT and OFDFT results of~\citet{Sjostrom_15} and kinetic theory-based results of~\citet{Daligault_16}. The values at low temperatures, from 1500~K (0.1293~eV) to 10000~K (0.8617~eV) are shown in the inset.}  
\label{fig:diff} 
\end{figure}

The values of the diffusion coefficient as a function of temperature are displayed in~\cref{fig:diff}. The diffusion coefficient increases with an increase in temperature, which is a direct consequence of the increased thermal motion of aluminum atoms at higher temperatures. We compared our predicted diffusion coefficients with DFT-MD and OFDFT calculations of~\citet{Sjostrom_15} and with kinetic theory calculations of~\citet{Daligault_16}. Our calculations follow the trend in and out of the training data range. Particularly at low temperature, our predictions agree well with accurate DFT-MD calculations. As temperature increases, our predictions start deviating somewhat from those reported in the literature. The reason for the discrepancy might be due to the use of different levels of theory. As temperature reaches 116040~K (10~eV), the difference between our prediction and OFDFT starts decreasing, likely due to the suitability of OFDFT at high temperatures.

Furthermore, we can check the consistency of our result by utilizing the Stokes-Einstein relation~\cite{Hansen_13} as an empirical reference result. It reads $D = k_B T/2 \pi d \eta$ and relates the shear viscosity with the diffusion coefficient, where $d$ denotes the diameter of a macroscopic sphere moving with constant velocity in a fluid of a given shear viscosity. Following~\citet{Alemany_04}, we apply this macroscopic concept to our case by determining the value of $d$ from the first peak in the radial distribution function $g(r)$ and using those viscosity values that were obtained from the Green-Kubo relation. At 10000 K, $d = 2.4975$ \AA\, and $\eta = 0.8317$ mPa$\cdot$s, the Stokes-Einstein relation yields as diffusion coefficient $1.0578 \times 10^{-4}$ $\text{mm}^2/\text{s}$ which is in good agreement with the calculation of the mean squared displacement 
yielding a value of $9.132 \times 10^{-5}$ $\text{mm}^2/\text{s}$.

\subsection{Longitudinal Collective Modes}

\subsubsection{Sound Velocity}
\begin{figure}[!hbt]   
\centering       
\includegraphics[width=1.0\columnwidth]{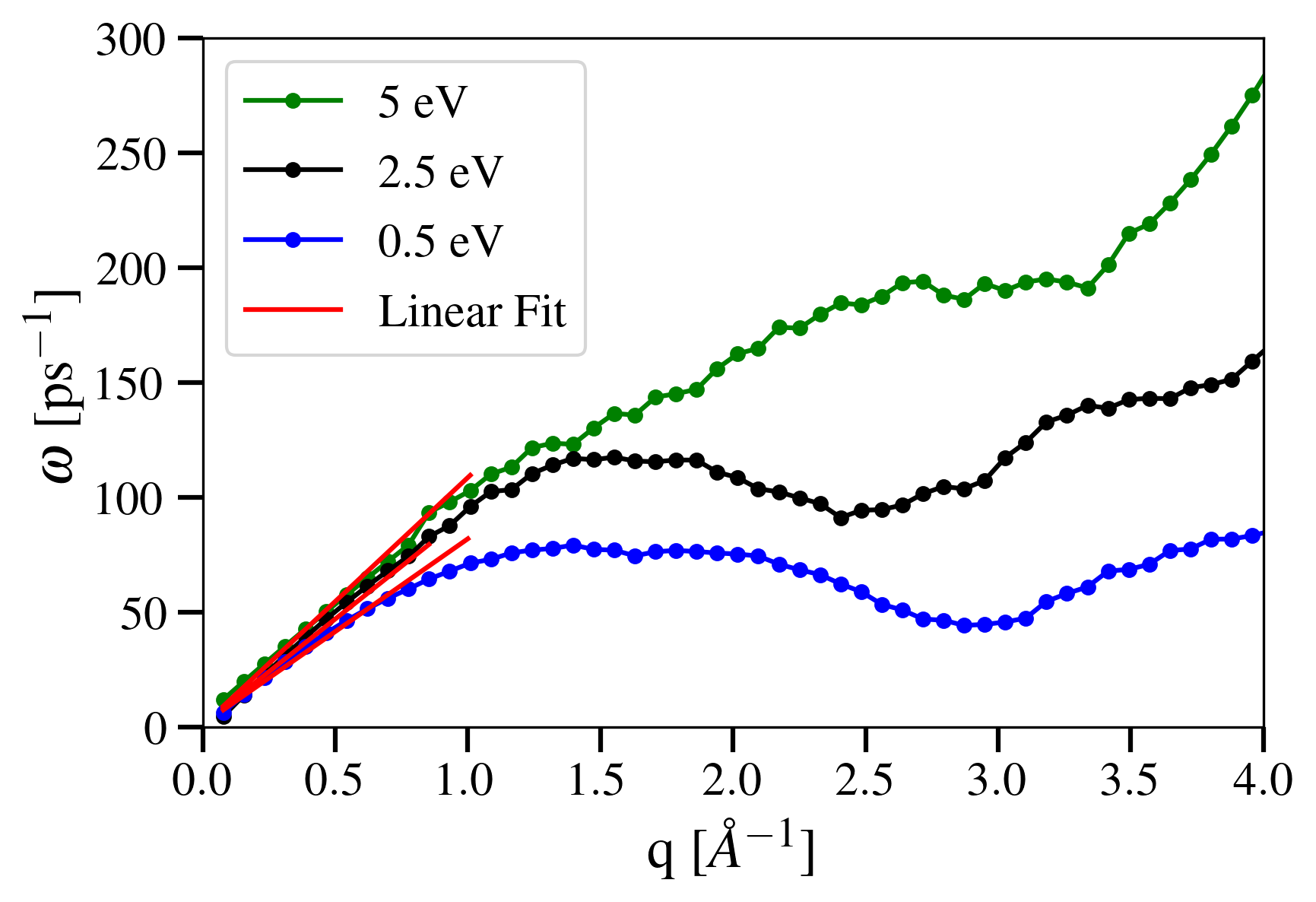}
\caption{Dispersion relation of aluminum at mass density of 2.7 $\text{g/cm}^3$ and temperature of 0.5~eV (5802~K), 2.5~eV (29011~K), and 5~eV (58022~K). The slope of the plot (red solid line) in the hydrodynamic regime (small wave vector q) provides sound velocity in the melted aluminum.}  
\label{fig:disp_rel} 
\end{figure}

The sound velocity of a material at a given temperature can be determined from the slope of the longitudinal dispersion relation $\omega(q)$ in the limit $q\to 0$~\cite{Hansen_75, Alemany_04}. Calculating $\omega(q)$ is computationally expensive because increasingly larger simulation cells are needed as $q$ decreases. Resolving the behavior of $\omega(q)$ at small $q$ becomes feasible in terms of an ML-IAP which enables us to calculate the dynamics in very large simulation cells and, hence, gather data for small $q$. 
The longitudinal current spectrum of ions is defined as 
\begin{equation}
\lambda(\mathbf{q},t)=\sum_{j}v_{jx}(t)e^{i\mathbf{q}.\mathbf{x}_{j}(t)}\,,
\end{equation}
where $q$ is the wave vector chosen along the x axis, which depends on integer number $n$ ($n$ = 1, 2, 3, 4, ...) and the simulation system length $L_x$. The longitudinal current correlation spectrum
\begin{equation}
L(\mathbf{q},\omega)=\frac{1}{2\pi N}\lim_{\tau \to \infty}\frac{1}{\tau}|\lambda(\mathbf{q},\omega)|^2,
\end{equation} 
is then calculated from the longitudinal current, where 
\begin{equation}
\lambda(\mathbf{q},\omega)=\int_{0}^{\tau}\lambda(\mathbf{q},t)e^{-i\omega t} dt\ .
\end{equation}
denotes the Fourier transform of $\lambda(\mathbf{q},t)$ and $\tau$ the simulation time which is truncated at a sufficiently large value in our calculations. The peaks in $L(\mathbf{q},\omega)$ correspond to the collective modes, which represent the maximum energy of the wave mode. The dispersion relation of the longitudinal wave mode can then be obtained by plotting $\omega(q)$ (the value at the peak) versus the wave vector $\mathbf{q}$, as shown in~\cref{fig:disp_rel}.

The sound velocity corresponds to the slope at small values for $q$ (linear region). For aluminum at temperatues of 0.5 eV, 2.5 eV, and 5 eV, we determine the sound velocity as 6450 m/s, 8887 m/s, and 9818 m/s, respectively. These values are in agreement with the sound velocities of 6273 m/s and 10365 m/s at 0.5 eV and 5 eV, respectively, obtained previously from OFDFT~\cite{White_13}.

\subsubsection{Ion-Ion Structure Factor}
The dynamic ion-ion structure factor (DSF) is obtained from density fluctuations~\cite{Alemany_04, Hansen_75} and is defined as 
\begin{equation}
S(\mathbf{q},\omega)= \frac{1}{2\pi}\int_{-\infty}^{\infty}F(\mathbf{q},t)e^{-i\omega t} dt
\end{equation}
in terms of the intermediate scattering function
\begin{equation}
F (\mathbf{q}, t) = \frac{1}{N} \langle n (\mathbf{q}, t) n(-\mathbf{q}, 0)\rangle
\end{equation}
from atomic density correlations $\langle n (\mathbf{q}, t) n(-\mathbf{q}, 0)\rangle$ obtained from an ensemble average. The atomic density is calculated from the atomic positions as 
\begin{equation}
n (\mathbf{q}, t) = \sum_{j} e^{i\mathbf{q}.\mathbf{x}_j(t)}\ .
\end{equation}  
Finally, the initial value of the intermediate scattering function yields the static structure factor (SSF)
\begin{equation}
S(\mathbf{q})= F (\mathbf{q}, t =0)\ . 
\end{equation}

Both the SSF and the DSF are measured in experiment, for example using neutron scattering or x-ray diffraction. Both are used for the diagnostics of the temperature, density, and ionization state in dense plasmas and WDM~\cite{Glenzer_09}. 
In this work, we compute the DSF and the SSF by post-processing our MD data with the DYNASOR code~\cite{Fransson_21}.

\begin{figure}[!hbt]   
\centering       
\includegraphics[width=1.0\columnwidth]{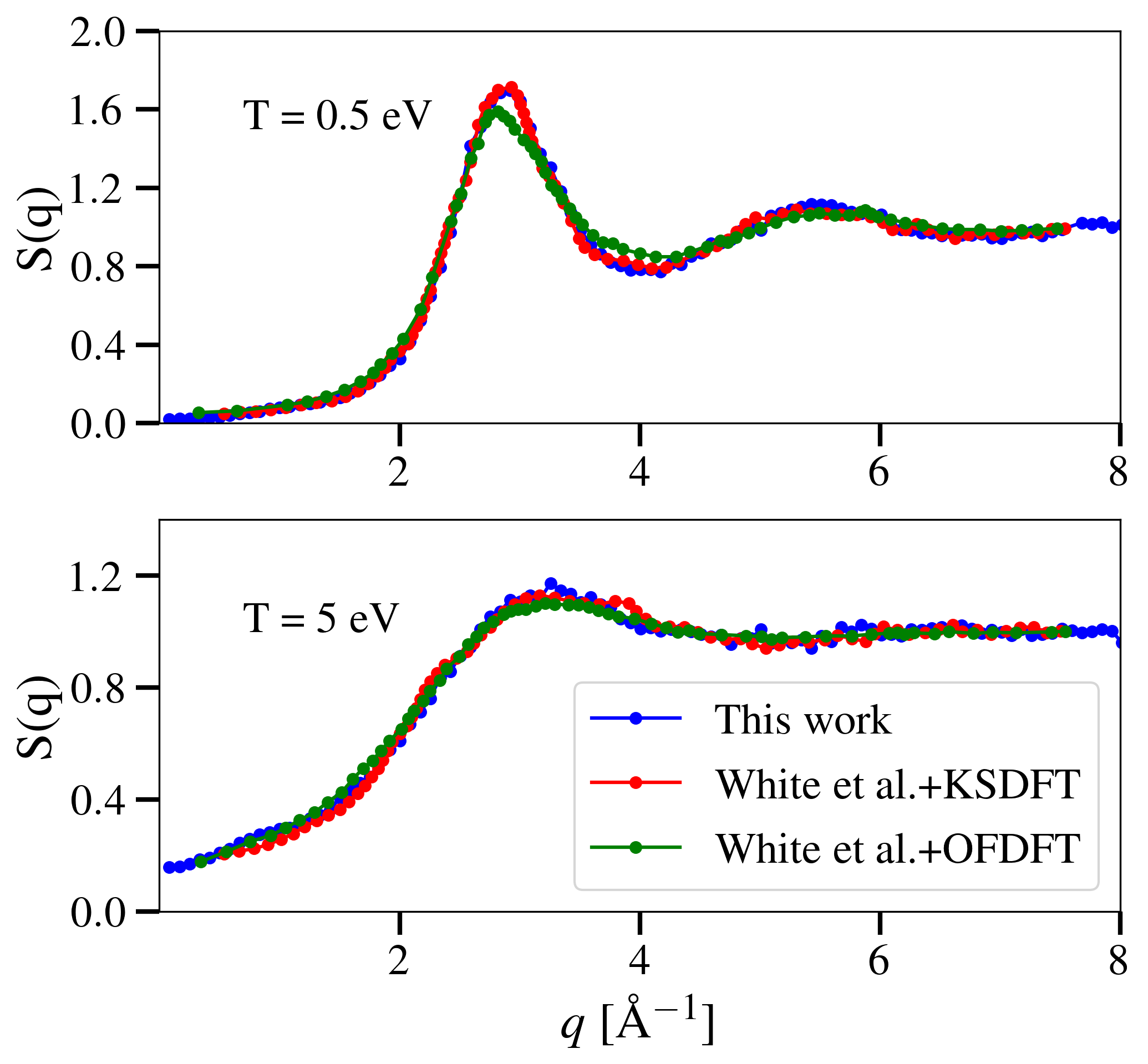}
\caption{Static structure factor of aluminum at mass density of 2.7 $\text{g/cm}^3$ and temperatures of 0.5~eV (5802~K) and 5~eV (58022~K). We compare our results (blue curve) with the DFT-MD (red curve) and OFDFT (green curve) results of~\citet{White_13}. Our results are in excellent agreement with the results obtained from DFT-MD.}  
\label{fig:ssf_plot} 
\end{figure}

First, we illustrate the SSF as a function of wave vector $q$ for temperatures of 0.5~eV (5802~K) and 5~eV (58022~K) in \cref{fig:ssf_plot} and compare our results with those from DFT-MD and OFDFT~\cite{White_13}. Note the excellent agreement of our prediction with the DFT-MD result at 0.5~eV (5802~K) and inaccuracies of the OFDFT result in the main peak. This confirms the abilities of our SNAP ML-IAP to interpolate accurately in the range of training data. Furthermore, we observe excellent agreement of our SNAP ML-IAP predictions with DFT-MD also at 5~eV (58022~K) indicating the reliability the SNAP ML-IAP also outside the range of training data.

\begin{figure}[!hbt]  
\centering       
\includegraphics[width=1.0\columnwidth]{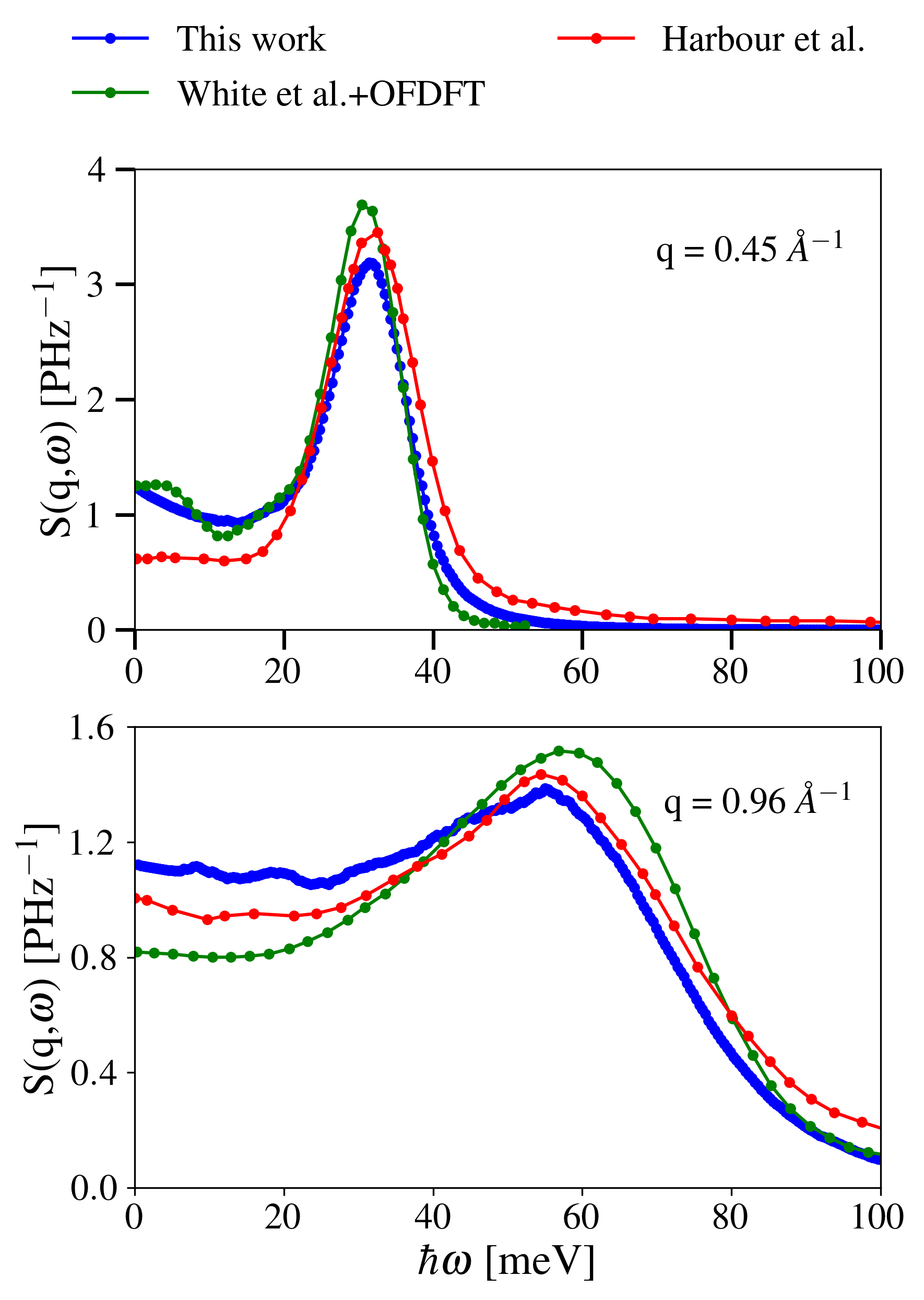}
\caption{Dynamic structure factor of aluminum at 2.7 $\text{g/cm}^3$ and 5~eV (58022~K) for wave vectors $\text{q}$ = 0.45 \AA $^{-1}$ (above) and $\text{q}$ = 0.96 \AA $^{-1}$ (below). Our result (blue) is compared with the OFDFT result of~\citet{White_13} (green curve) and the neutral pseudoatom (NPA) model of~\citet{Harbour_18} (red curve).} 
\label{fig:dsf_plot} 
\end{figure}

Finally, we illustrate the DSF at 5~eV (58022~K) for two values of the wave vector $\text{q}= 0.45$ \AA$^{-1}$ and $\text{q}$ = 0.96 \AA $^{-1}$ in~\cref{fig:dsf_plot}. Similarly, we compare our results with result from OFDFT~\cite {White_13} (green curve) and a neutral pseudoatom (NPA) model ~\cite{Harbour_18} (red curve) and find qualitative agreement. We can further assess the quality of our predictions by checking sum rules the DSF obeys~\cite{Wax_13}. To that end, we evaluate the sum rule
\begin{align}
\int_{-\infty}^{\infty} d\omega S(\mathbf{q}, \omega) = S(\mathbf{q})= F (\mathbf{q}, 0)
\end{align}
which states that the SSF is recovered from the DSF by integration over the frequency domain. This is demonstrated for temperatures of 0.5~eV (5802~K) and 5~eV (58022~K) in Fig. S4~\cite{sup_mat}.

%
%

\section{Conclusions} 
\label{sec:concl}
Our study has resulted in the training of a single ML-IAP using DFT-MD data and the SNAP methodology to model the properties of aluminum in a wide range of conditions, from ambient to WDM. Previous ML-IAPs have been trained for specific temperature and pressure ranges. However, our findings demonstrate that a single ML-IAP can yield reliable and precise results across a broad temperature range.

To evaluate the transferability of our ML-IAP, we computed several materials and transport properties, including the thermal conductivity, viscosity, diffusion coefficient, sound velocity, and ion-ion structure factor. 
Our results for the thermal conductivity and viscosity are consistent with previous findings from ML-IAPs~\cite{Zhou_07, Stojanovic_10, Liu_21, Cheng_21}, although those ML-IAPs were specifically trained for the tested temperature ranges. Regarding the diffusion coefficient, our predictions agree well with reference data from DFT-MD~\cite{Sjostrom_15} and are in qualitative agreement with less accurate results from OFDFT~\cite{Sjostrom_15} and kinetic theory~\cite{Daligault_16}. Furthermore, our ML-IAP enables the determination of the sound velocity, which is otherwise difficult to compute using DFT-MD calculations due to the requirement of large simulation cells. We demonstrate these calculations for aluminum at a broad temperature range and obtain qualitative agreement with OFDFT results~\cite{White_13}. Lastly, we show that our ML-IAP can predict the ion-ion structure factor with the same accuracy as DFT-MD~\cite{White_13}.  
Overall, our results demonstrate the ability of our ML-IAP to yield results in line with prior research and, more importantly, its reliability and accuracy beyond the training range. 

Demonstrating the transferability of an ML-IAP across a broad temperature range presents a significant advancement. Utilizing transferable ML-IAPs reduces the computational resources required to generate materials properties for studying matter under elevated temperatures and enables more comprehensive and accurate simulations of materials properties. This will greatly support the interpretation of experimental measurements of laser-driven and shock-compressed samples at free-electron laser facilities worldwide. While we have shown the transferability of our ML-IAP for temperature ranges, a comprehensive ML-IAP that is useful for extended simulations of properties in WDM requires transferability across both pressure and temperature ranges. We plan to address this challenge in our future work and further advance our methodology.

\begin{acknowledgments} 
This work was partially supported by the Center for Advanced Systems Understanding (CASUS) which is financed by Germany’s Federal Ministry of Education and Research (BMBF) and by the Saxon state government out of the State budget approved by the Saxon State Parliament. ML was supported by the German Federal Ministry of Education and Research (BMBF, No. 01/S18026A-F) by funding the competence center for Big Data and AI “ScaDS.AI Dresden/Leipzig.”  Computations were performed on a Bull Cluster at the Center for Information Services and High Performance Computing (ZIH) at Technische Universit\"at Dresden, on the cluster Hemera at Helmholtz-Zentrum Dresden-Rossendorf (HZDR). This article has been authored by an employee of National Technology \& Engineering Solutions of Sandia, LLC under Contract No. DE-NA0003525 with the U.S. Department of Energy (DOE). The employee owns all right, title and interest in and to the article and is solely responsible for its contents. The United States Government retains and the publisher, by accepting the article for publication, acknowledges that the United States Government retains a non-exclusive, paid-up, irrevocable, world-wide license to publish or reproduce the published form of this article or allow others to do so, for United States Government purposes. The DOE will provide public access to these results of federally sponsored research in accordance with the DOE Public Access Plan https://www.energy.gov/downloads/doe-public-access-plan.
\end{acknowledgments}







%
 
\end{document}


\newpage
\tableofcontents
\clearpage

\section{Training data sets, energy weights, and force weights}
We provide DFT-MD data sets used for the training of spectral Neighbor Analysis Potential (SNAP) in \cref{tab:snap_para}. Dakota optimized energy weight (EW) and force weight (FW) for SNAP potential are also provided in \cref{tab:snap_para} for each configuration in respective order.  

\begin{table}[!hbt]
\caption{The training data sets used for SNAP ML-IAP training are provided along with the optimized energy weights (EWs) and force weight (FWs) for each configuration in respective order.}
\begin{center}
\begin{tabular}{ccccc}
\hline\hline
T [K] & No. Atoms & No. Configs. & EW [$10^{-3}$] & FW [$10^{-5}$]\\
\hline
300 & 128, 256 & 97, 99 & 14, 8.76  & 2.04, 1.05 \\
400 & 128, 256 & 97,99 & 11.5, 7.42  & 2.05, 9.22 \\
500 & 128, 256 & 98, 100 & 6.28, 8.69  & 3.86, 1.86 \\
600 & 128, 256 & 98, 100 & 14.34, 9.58  & 2.21, 0.72 \\
700 & 128, 256 & 98, 100 & 7.33, 4.39  &  1.16, 1.33 \\
800 & 128 & 97 & 5.93 & 3.05 \\
900 & 128, 256 & 96, 99 & 10.26, 12.05 & 3.19, 1.67 \\
1000 & 128, 256 & 93, 100 & 11.47, 6.71 & 1.75, 0.78 \\
2000 & 128, 256 & 93, 100 & 9.868, 11.04  & 2.718, 1.597\\
3000 & 128 & 97 & 6.44 & 2.503 \\
4000 & 128, 256 & 97, 99 & 5.646, 10.37  & 2.743, 1.284\\
5000 & 128, 256 & 92, 99 & 5.04, 10.72  & 1.641, 1.418\\
6000 & 128, 256 & 96, 100 & 9.539, 14.69  & 3.27, 0.888\\
7000 & 128, 256 & 98, 99 & 12.34, 1.478  &  4.35, 1.82\\
8000 & 128, 256 & 93, 100 & 6.479, 8.054  & 1.887, 0.8297\\
9000 & 128, 256 & 98, 100 & 7.423, 9.134  & 3.569, 1.157\\
10000 & 128, 256 & 98, 99 & 9.375, 1.054  & 3.907, 0.71\\
\hline\hline
\end{tabular}
\end{center}
\label{tab:snap_para}
\end{table} 

\clearpage

\section{Assessing the accuracy of the SNAP ML-IAP}
\begin{figure}[!hbt]   
\centering       
\includegraphics[width=0.6\columnwidth]{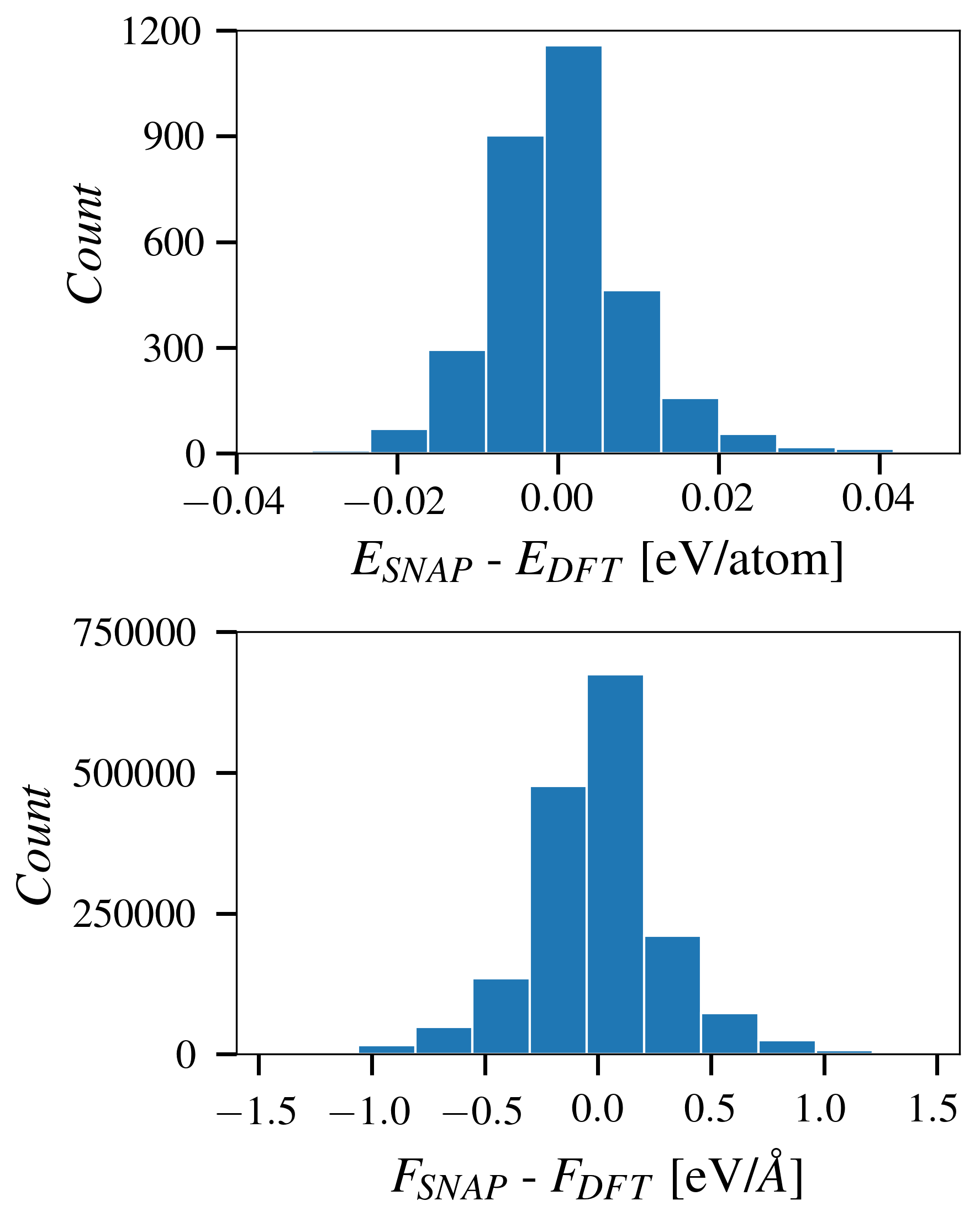}
\caption{Histogram showing error of the SNAP ML-IAP for energy and force predictions with respect to the reference DFT-MD data.}
\label{fig:disp_rel} 
\end{figure}

\begin{figure}[!hbt]
\centering   
\includegraphics[width=0.6\columnwidth]{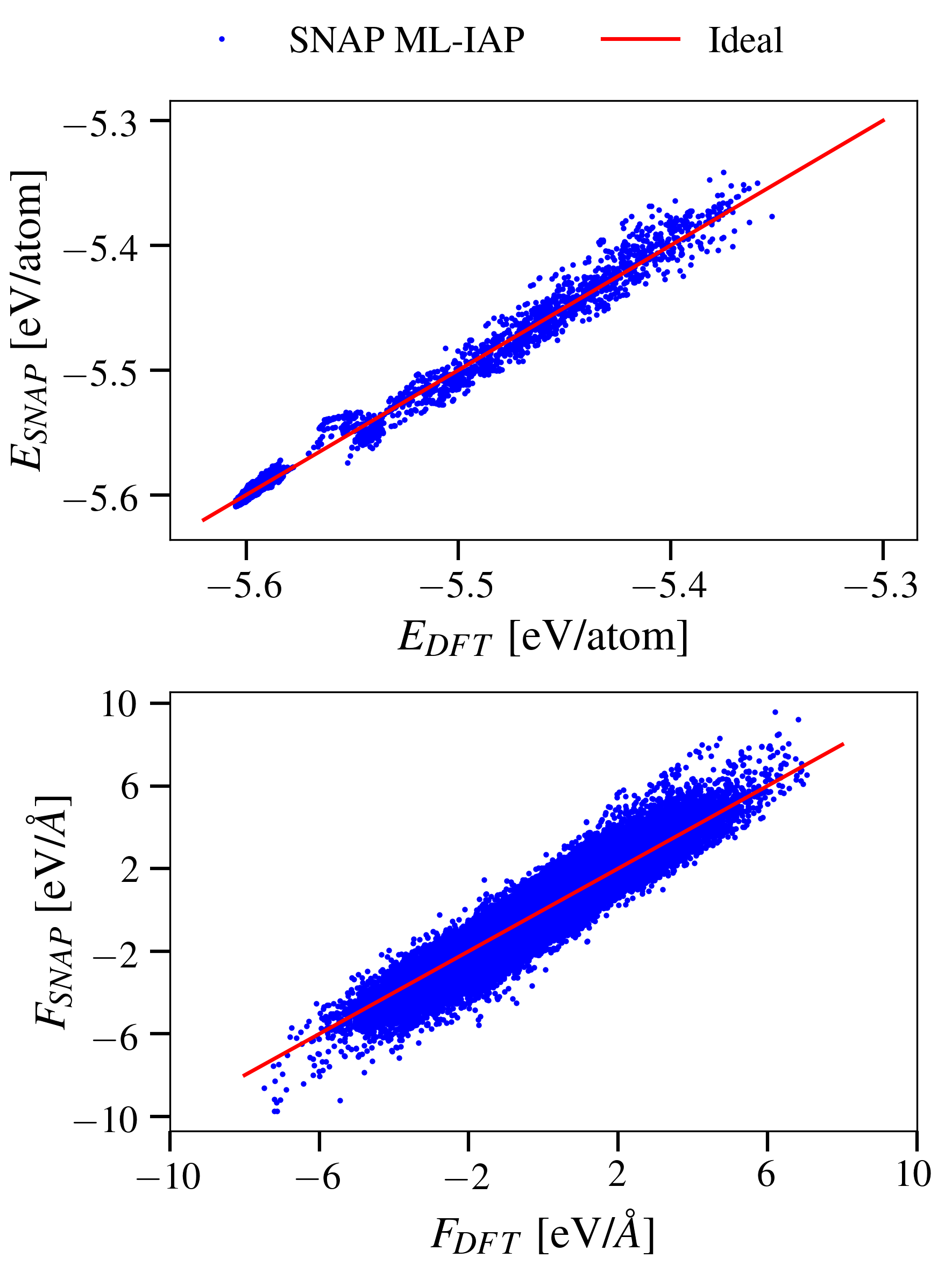} 
\caption{Correlation plot of the energy and force predictions illustrating the qualitative accuracy of the SNAP ML-IAP relative to the DFT-MD reference data.} 
\label{fig:Error_fit_energy}   
\end{figure}

We plot histograms that illustrate the error of the SNAP machine-learned interatomic potential (ML-IAP) energy and force predictions with respect to the DFT-MD reference values. As shown in \cref{fig:disp_rel}, the distribution of the errors is centered around zero. Furthermore, we assess the accuracy of our SNAP ML-IAP in terms of a correlation plot for both energies and forces in \cref{fig:Error_fit_energy}. The spread of SNAP ML-IAP predictions around the ideal diagonal provides a qualitative measure of accuracy. The shown results indicate that the trained potential is suitable for the MD simulations. 


\clearpage 
\section{Assessing finite-size effect in the thermal conductivity}
\begin{figure}[!hbt]   
\centering       
\includegraphics[width=0.6\columnwidth]{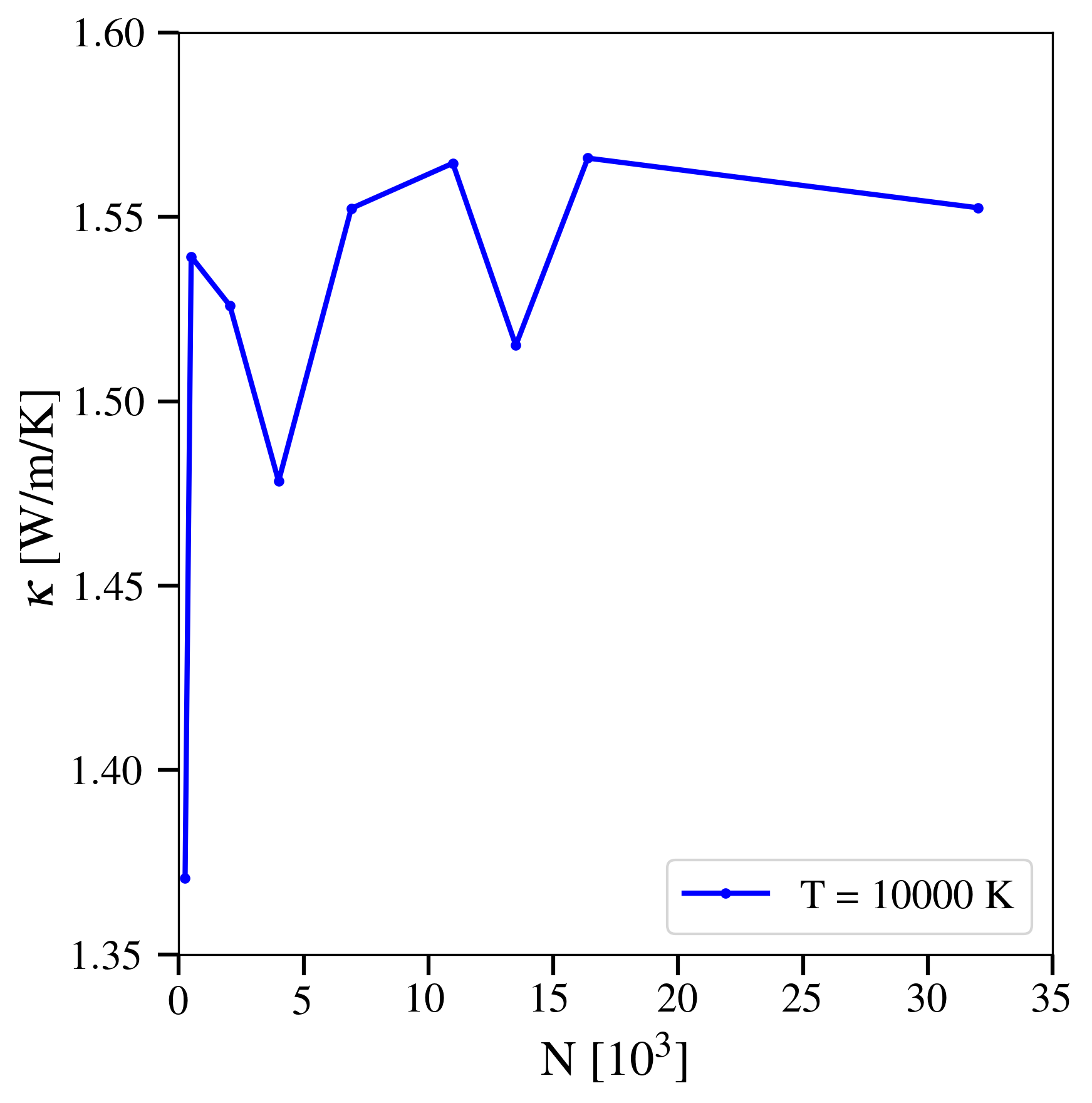}
\caption{Lattice thermal conductivity as a function of the number of atoms.}  
\label{fig:thc_finite} 
\end{figure}

We assess the finite-size effects on the thermal conductivity in \cref{fig:thc_finite} by performing MD simulations with an increasing atom count. Beyond an atom count of 5000, the fluctuation in the thermal conductivity is around 3$\%$. Our analysis reveals that a number of 13500 atoms is needed to achieve converged results and suppress finite size errors. 

\section{Assessing the dynamic structure factor in terms of sum rules}
\begin{figure}[!hbt]   
\centering       
\includegraphics[width=0.7\columnwidth]{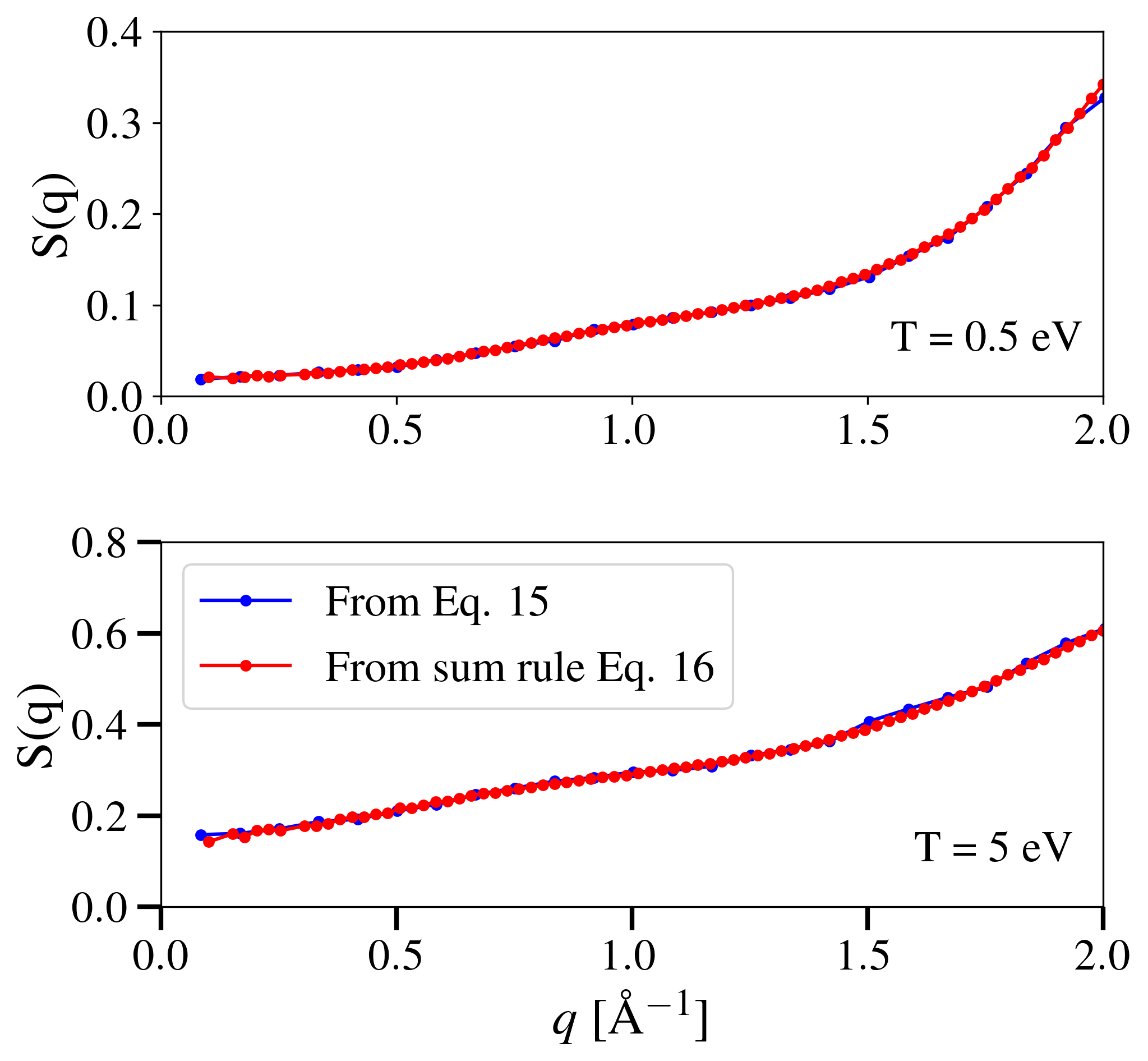}
\caption{Static structure factor at temperatures of 0.5~eV (5802~K) and 5~eV (58022~K) from in the intermediate scattering function [Eq.~(15) of our manuscript] and from the sum rule [Eq.~(16) of our manuscript].}  
\label{fig:comp_sum} 
\end{figure}

We assess the quality of the predicted dynamic structure factors in terms of the sum rules. In \cref{fig:comp_sum}, we compare the static structure factors for temperatures of 0.5~eV (5802~K) and 5~eV (58022~K) computed from the sum rule  [Eq.~(16) of our manuscript] with those obtained directly from the intermediate scattering function [Eq.~(15) of our manuscript]. It is evident that the dynamic structure factor obeys the sum rule indicating the reliability of our calculations and the quality of our ML-IAP.

\clearpage 
%
%
%
%
